\documentclass[reprint,aps,floats,floatfix,amsmath,amssymb,amsfonts,nofootinbib,longbibliography,superscriptaddress]{revtex4-1}

\usepackage{graphicx}
\usepackage{dcolumn}
\usepackage{bm}
\usepackage{tensor} 
\usepackage{natbib}
\usepackage{subcaption}

\usepackage[hyperindex,colorlinks]{hyperref}

\begin{document}

\title{Nonconservative unimodular gravity: Gravitational waves}

\author{J\'ulio C. Fabris}
\email{julio.fabris@cosmo-ufes.org}%
\affiliation{%
N\'ucleo Cosmo-ufes \& Departamento de F\'isica,  Universidade Federal do Esp\'irito Santo (UFES)\\
Av. Fernando Ferrari, 540, CEP 29.075-910, Vit\'oria, ES, Brazil.}%
\affiliation{%
National Research Nuclear University MEPhI, Kashirskoe sh. 31, Moscow 115409, Russia}%

\author{Marcelo H. Alvarenga}
\email{marcelo.alvarenga@edu.ufes.br}
\affiliation{%
N\'ucleo Cosmo-ufes \& Departamento de F\'isica,  Universidade Federal do Esp\'irito Santo (UFES)\\
Av. Fernando Ferrari, 540, CEP 29.075-910, Vit\'oria, ES, Brazil.}%
\author{Mahamadou Hamani Daouda}%
\email{daoudah77@gmail.com}

\affiliation{Département de Physique - Université de Niamey, Niamey, Niger.}%

\author{Hermano Velten}%
\email{hermano.velten@ufop.edu.br}
\affiliation{%
Departamento de F\'isica, Universidade Federal de Ouro Preto (UFOP), Campus Universit\'ario Morro do Cruzeiro, 35.400-000, Ouro Preto, Brazil}%
\date{\today}

\begin{abstract}

Unimodular gravity is characterized by an extra condition with respect to General Relativity: the determinant of the metric is constant. This extra condition leads to a more restricted class of invariance by
coordinate transformation. Even though, if the conservation of the energy-momentum tensor is imposed in unimodular gravity, the General Relativity theory is recovered with an additional integration constant which is associated to the cosmological term $\Lambda$. However, if the energy-momentum tensor does not conserve separately, a new geometric structure appears with potentially observational  signatures. In this text, we consider the evolution of gravitational waves in the nonconservative unimodular gravity, showing how it differs from the usual signatures 
in the standard model. 
\end{abstract}

\maketitle

\section{Introduction}

General Relativity (GR) is considered as the modern theory of gravitation since its final formulation in 1915. However, soon after the GR theory has been introduced many other theories for the gravitational interaction have been proposed. Unimodular gravity is one of the oldest alternative to GR, see for example Ref. \cite{wein-a} for a discussion on this alternative formulation of a gravity theory. Unimodular gravity is similar to GR in many respects. One important similarity is that both are geometric theories of gravity based on a Lagrangian containing the Ricci scalar. However, unimodular gravity has one special new ingredient: the determinant of the metric is imposed to be a constant which can be set equal to 1. In practise, this leads to equations that are traceless. Then, source of the geometrical structure, i.e., the right hand side of the field equations, must also correspond to a traceless structure.

In GR theory, the Bianchi identities lead automatically to conservation of the energy-momentum tensor. This is connected to the invariance of the theory by general diffeomorphisms. The unimodular gravity has a more restricted class of transformation due to the condition on the determinant of the metric, usually called transverse diffeomorphism. For a detailed
discussion on the transverse diffeomorphism
see for example Ref. \cite{trans} and references therein. The invariance by this restricted class of transformations, with respect to the general diffeomorphism, in the unimodular gravity has one particular consequence for the divergence of the energy-momentum tensor: it can be zero, implying the usual conservation laws, and the GR equations are recovered with an extra integration constant which can be identified with the cosmological constant; or the energy-momentum tensor does not conserve, and a radical new structure appears. In this article, we want to explore the second option as described later.

The unimodular condition on the determinant of the metric implies that volumes must be preserved in a coordinate transformation. This property, together with the other aspects related to the notion of transverse diffeomorphism, has been explored in many studies related to quantum gravity
\cite{quantum1}. See also Ref. \cite{Herrero-Valea:2018ilg} whose results are more relevant to our work.

The unimodular equations are traceless but not conformal invariant. Traceless equations in gravity theory have been evoked
many times in the literature, see for example Ref. \cite{ellis}. In Ref. \cite{mimetic1,mimetic2} the mimetic gravity theory has been proposed, which may be written also in a traceless form, and which have many interesting consequences for the early and late cosmological scenarios.
Here we will restrict ourselves to the unimodular gravity. In general, in the investigations involving unimodular gravity, the conservation of the energy-momentum tensor is imposed. This imposition leads to some nice properties. One of them is that the theory becomes equivalent to GR in presence of a cosmological term that appears an integration constant. In some sense the diffeomorphism invariance may be recovered, see for example the discussion in Ref. \cite{wein-a}. 

However, the second option described above, the non conservation of the energy-momentum tensor, may open many interesting possibilities. One of them is that the resulting cosmological model is equivalent, at the background level, to cosmological model in GR with only radiation and a cosmological constant. This can be seen as a bad feature, since in the standard cosmological model based in the GR theory a matter dominated phase is necessary in order to allow for a successful large structure formation process. But, in the unimodular gravity the perturbative features may allow structure formation even in a radiative universe. This problem is treated more in detail in a companion paper \cite{marcelo}.

In the present paper we have a very specific goal: to study the evolution of gravitational waves in a cosmological model
resulting from unimodular gravity with non conservation of the energy-momentum tensor. We will call it nonconservative unimodular gravity. 
We will verify that gravitational waves in this nonconservative context have specific features that may allow for testing the model with future observational data on primordial gravitational waves.
Our analysis will be fundamentally qualitative for reasons to be discussed later in this article, but opens possibilities for a more quantitative study.

The paper is organized as follows. In the next section we outline the main features of the unimodular gravity. In section III we discusss the construction of the nonconservative unimodular gravity. In section IV we compute the evolution of the cosmological gravitational waves in the nonconservative unimodular gravity and in section V we compare it with know results in the standard cosmological model.
In section VI we discuss our main results and present our conclusions.

\section{Unimodular gravity}

In order to better evaluate the meaning of the unimodular condition let us briefly first review aspects of the structure of GR theory. Here we follow closely a similar exposition made in Ref. \cite{brand}.

The equations of GR are obtained through variational principle from the Einstein-Hilbert action,
\begin{eqnarray}
\label{gr-a}
{\cal S} = \int d^4x\sqrt{-g}\{R - 2\Lambda + {\cal L}_m\},
\end{eqnarray}
where ${\cal L}_ m$ denotes the matter Lagrangian density, $\Lambda$ is the cosmological constant and $R$ is the Ricci scalar. This is the most simple action allowed in four dimensions leading to second order differential equations.
We have already introduced $\Lambda$ in the action since it is allowed by the Lovelock theorem. The presence of the
cosmological term may also allow to implement later the standard cosmological model for the universe, the $\Lambda$CDM model.

The variation of the action (\ref{gr-a}) with respect to the metric $g_{\mu\nu}$ leads to the GR equations in presence of
matter and a cosmological constant:
\begin{eqnarray}
\label{rg-eq1}
R_{\mu\nu} - \frac{1}{2}g_{\mu\nu}R = 8\pi GT_{\mu\nu} + g_{\mu\nu}\Lambda.
\end{eqnarray}
Here, the energy-momentum tensor is defined as,
\begin{eqnarray}
T_{\mu\nu} = -\frac{2}{\sqrt{-g}}\frac{\delta \sqrt{-g}{\cal L}_m}{\delta g^{\mu\nu}}.
\end{eqnarray}
The use of the Bianchi identities lead the conservation of the energy-momentum tensor,
\begin{eqnarray}
\label{rg-eq2}
{T^{\mu\nu}}_{;\mu} = 0.
\end{eqnarray}
This property is directly connect with the invariance of the GR theory with respect to 
difeomorphism, see Ref. \cite{wald}.

Now we turn to unimodular gravity. In principle, we must follow the same procedure used to obtain the GR equations, but implementing at the same time the unimodular condition, $g = $ constant (constant that may be equal to 1).
In general, the introduction of a constraint in an action can be made with the help of Lagrange's multiplier. In doing so, in the unimodular case, some details must be outlined. Let us consider, for example, the unimodular condition for a flat, isotropic and homogenous universe, described by the metric,
\begin{eqnarray}
ds^2 = N^2 dt^2- a(t)^2(dx^2 + dy^2 + dz^2),
\end{eqnarray}
where $N$ is the lapse function and $a(t)$ is the scale factor. In this case, the unimodular condition, with the constant equal to one, implies,
\begin{eqnarray}
Na^3 = 1.
\end{eqnarray}
Hence, the unimodular condition would fix the lapse function, $N = a^{-3}$. We will see below how to circumvent this limitation.

In order to construct the unimodular equations we use, as already stated, the Lagrangian multipliers in the action:
The action reads,
\begin{eqnarray}
\label{uni-a1}
{\cal S} = \int d^4x\biggr\{\sqrt{-g}R - \chi(\sqrt{-g} - 1)\biggl\} + \int d^4 x \sqrt{-g}{\cal L}_m,
\end{eqnarray}
where all the quantities are defined as in GR, and $\chi$ is the Lagrangian multiplier.
To circumvent the restriction on the lapse function described above, we modify the action (\ref{uni-a1}), rewriting it as,
\begin{eqnarray}
\label{uni-a2}
{\cal S} = \int d^4x\biggr\{\sqrt{-g}R - \chi(\sqrt{-g} - \xi)\biggl\} + \int d^4 x \sqrt{-g}{\cal L}_m.
\end{eqnarray}
where $\xi = \xi(t)$ is an arbitrary function of time which is not subjected to the variational principle: $\xi(t)$ may be considered as an external field.

Varying the action (\ref{uni-a2}) with respect to the metric, it follows,
\begin{eqnarray}
\label{erg1}
R_{\mu\nu} - \frac{1}{2}g_{\mu\nu}R + \frac{\chi}{2}g_{\mu\nu} = 8\pi GT_{\mu\nu}.
\end{eqnarray}
Varying with respect to $\chi$ we obtain the unimodular constraint.
\begin{eqnarray}
\label{vin-rg-1}
\xi = \sqrt{-g}.
\end{eqnarray}
Since, $\xi$ is an arbitrary function of time, now any choice for the lapse function is possible, as in GR.

From the trace of (\ref{erg1}) we obtain
\begin{eqnarray}
\chi = \frac{R}{2} + 8\pi G\frac{T}{2}.
\end{eqnarray}
Inserting this result in (\ref{erg1}) we obtain,
 \begin{eqnarray}
 \label{erg2}
 R_{\mu\nu} - \frac{1}{4}g_{\mu\nu}R = 8\pi G\biggr(T_{\mu\nu} - \frac{1}{4}g_{\mu\nu}T\biggl).
 \end{eqnarray}
 These are the unimodular field equations. Remark that the equations are traceless.
 
 Now we inspect the conservation laws. The use of the Bianchi identities lead to the following new conservation law:
 \begin{eqnarray}
 \label{brg1}
 \frac{R^{;\nu}}{4} = 8\pi G\biggr({T^{\mu\nu}}_{;\mu} - \frac{1}{4}T^{;\nu}\biggl).
 \end{eqnarray}
In principle, the GR conservation laws are not recovered. This is expected since the general diffeomorphism invariance is broken by the unimodular condition. But, it can be imposed as a new choice. In fact, if the relation ${T^{\mu\nu}}_{;\mu} = 0$ is imposed, we obtain the following expression:
 \begin{eqnarray}
 R^{;\nu} = - 8\pi G T^{;\nu}.
 \end{eqnarray}
 which can be integrated, leading to,
 \begin{eqnarray}
 R = 8\pi G T - \Lambda,
 \end{eqnarray}
 where $\Lambda$ is an integration constant which is identified with the cosmological term. The choice that the energy-momentum tensor conserves separately leads to the GR equation with a cosmological term, that is,
 equations, (\ref{rg-eq1},\ref{rg-eq2}). On the other hand, if the conservation of the energy-momentum tensor is not imposed, we have the field equations given (\ref{erg2}) with the {\it generalized} conservation law (\ref{brg1}).
 It is these last set of equations that we want to study here, considering the evolution of gravitational waves in a cosmological context.
 
 \section{A Cosmological model}
 
Our next step in analyzing the nonconservative unimodular gravity theory is to explore the evolution of gravitational waves in the cosmological context. Cosmological models in unimodular gravity with a conserved energy-momentum tensor has been discussed, for example, in Ref. \cite{alvarez}. The present work will be concentrate in this specific problem which may already give some insight on
the consequences of using equations (\ref{erg2}) and (\ref{brg1}). In these equations we introduce the flat metric, with the lapse function given by $N = 1$, a possibility allowed as discussed in the previous section. Hence, the line element is given by,
\begin{eqnarray}
\label{metric}
ds^2 = dt^2 - a(t)^2(dx^2 + dy^2 + dz^2).
\end{eqnarray}
What is remarkable is that the time-time and space-space components of (\ref{erg2}) lead to the same equation:
\begin{eqnarray}
\label{ce1}
\dot H = - 4\pi G(\rho + p),
\end{eqnarray} 
where we have noted $H = \dot a/a$. It is worth noting that the above equation implies that the vacuum case corresponds to the de Sitter solution in contrast to the GR case where it corresponds to the Minkowski solution \cite{alvarez}.
This fact indicates that
the system of equations is underdetermined. 

But we still have the conservation equation (\ref{brg1}), which we can explore in order to a complete set of self-consistent equations.
The conservation laws result in the following equation:
\begin{eqnarray}
\label{ce2}
\ddot H + 4H\dot H &=& - 4\pi G[\dot\rho + \dot p + 4H(\rho + p)],
\end{eqnarray}
But both equations (\ref{ce1}) and (\ref{ce2}) surprisingly have the same content: inserting (\ref{ce1}) into (\ref{ce2}) we obtain $0 = 0$. Even with the use of the generalized conservation law, the system of equations remain underdetermined.

At this stage, two observations are in order: as already stated, the resulting system of equations in the cosmological context (with isotropy and homogeneity) is underdetermined, since there are two functions, $\rho$ and $H$, for just one equation, remembering that the pressure is connected to density by an equation of state; moreover, equations (\ref{ce1}) and (\ref{ce2}) are sensitive only
to the combination $\rho + p$, which is related to the enthalpy of the system, and does not depend on the equation of state provided it is barotropic, $p = p(\rho)$.

Defining $\bar \rho = \rho + p$, the equations (\ref{ce1},\ref{ce2}) can be written as,
\begin{eqnarray}
\label{ce3}
\dot H &=& - 4\pi G\bar\rho,\\
\label{ce4}
\ddot H + 4H\dot H &=& - 4\pi G(\dot{\bar\rho} + 4H\bar\rho), 
\end{eqnarray}
simplifying the notation.

An interesting aspect is that the usual radiative solution of GR 
\begin{eqnarray}
H = \frac{1}{2t}, \quad \bar\rho = \bar\rho_0 a^{-4},
\end{eqnarray}
is also solution for (\ref{ce3}) and (\ref{ce4}) irrespective of the pressure $p$, except for the case the pressure represents the vacuum energy or, alternatively, a cosmological term: $p \neq - \rho$. 

The fact that the radiative solution is also a solution of the nonconservative unimodular equations may be related to the traceless property of the field equations. Remark, however, that these field equations are traceless but not conformal invariant.
For $p = - \rho$ we find the usual de Sitter solution, $a \propto e^{\kappa t}$, $\kappa$ a constant (positive or negative).
Is there any other solution? Since the system is underdetermined, we can not obtain any other solution unless some additional hypothesis is introduced. We will later provide more comments on this point.

To obtain a specific model, and taking into account the properties discussed above, we can impose that both sides of (\ref{ce4}) conserves separately. This leads to,
\begin{eqnarray}
\label{cea}
\ddot H + 4H\dot H = 0,\\
\label{ceb}
\dot{\bar\rho} + 4H\bar\rho = 0.
\end{eqnarray}

The equation (\ref{ceb}) has a simple solution:
\begin{eqnarray}
\bar\rho = \bar\rho_0 a^{-4}.
\end{eqnarray}
This is typical of a radiative fluid. However, we remember that $\bar\rho = (\rho + p)$, hence the radiative behavior is independent of the fluid considered. This is reflection of the traceless character of the field equations and the nonconservation of the energy-momentum tensor (otherwise we go back to GR structure).

Besides the behavior for any fluid, which is always similar to a radiative fluid, independent of the corresponding pressure, there is another difference. In general relativity, the equations for a radiative fluid are given by,
\begin{eqnarray}
H^2 &=& \frac{8\pi G}{3}\rho,\\
2\dot H + 3H^2 &=& - \frac{8\pi G}{3}\rho.
\end{eqnarray}
These equations lead to,
\begin{eqnarray}
\dot H + 2H^2 = 0.
\end{eqnarray}
We notice that equation (\ref{cea}) can be written as
\begin{eqnarray}
\frac{d}{dt}\biggr(\dot H + 2H^2\biggl) = 0.
\end{eqnarray}
This is equivalent to the ansatz used in Ref. \cite{trace}, $R = $ constant which, in a similar context, has been introduced to complete the set of equations. The GR case corresponds to set the constant equal to zero. 

Following the reasoning exposed above, in the unimodular cosmology, without a separated conservation of the energy-momentum tensor, the equation for the 
Hubble function is given by,
\begin{eqnarray}
\label{cec}
\dot H + 2H^2 = \frac{2}{3}\Lambda_{\rm U}, \quad \Lambda_{\rm U} = \mbox{constant}.
\end{eqnarray}
The integration constant, which we have called $\Lambda_{\rm U}$ to keep the contact with the cosmological term, makes the unimodular cosmological scenario similar to the GR radiative model in presence of a cosmological constant, even if no cosmological constant was present from the beginning. If fact an integration constant similar to the cosmological constant is {\it hidden} in the structure of unimodular gravity.

From (\ref{cec}) we have three possibilities.
\begin{eqnarray}
\Lambda_{\rm U} < 0 \quad &\rightarrow& \quad a = a_0\sin^{1/2}\sqrt{- \frac{4\Lambda_{\rm U}}{3}}t,\label{sol1}\\
\Lambda_{\rm U} = 0 \quad &\rightarrow& \quad a = a_0 t^{1/2},\label{sol2}\\
\Lambda_{\rm U} > 0 \quad &\rightarrow& \quad a = a_0\sinh^{1/2}\sqrt{\frac{4\Lambda_{\rm U}}{3}}t.
\label{sol3}
\end{eqnarray}
These are essentially the same solutions found in Ref. \cite{trace}.
The case $\Lambda_{\rm U} = 0$ is identical to the GR radiative model. If $\Lambda_{\rm U} \neq 0$ we obtain formally the cosmological solutions in GR for a mixture of radiation and cosmological constant. The solutions corresponding to $\Lambda_{\rm U} \neq 0$ could also be expressed in terms of $\cos$ and $\cosh$ functions, which represent a nonsingular universe. But these possibilities would imply negative energy density $\bar\rho$, which mounts to a violation of the null energy condition since $\bar\rho$ leads to $\rho + p < 0$, and for this reason we avoid this possibility.

For all three values of $\Lambda$ the behavior for $t \rightarrow 0$ are similar, and coincides with the flat radiative case. 
Of particular interest is the case $\Lambda_{\rm U} > 0$: it interpolates the initial radiative phase and a de Sitter phase. In the usual GR context such possibility is generally not explored since structure formation requires a matter dominated phase ($p = 0$), which would be here absent. But such restriction may be circumvented in the nonconservative unimodular scenario
as it is discussed in \cite{marcelo}. The case $\Lambda_{\rm U} < 0$ corresponds to an interpolation between a radiative phase and an anti-de Sitter phase which in practise is not interesting from the observational point of view.

\section{Gravitational waves}

We will perform now the perturbative computation of the tensorial modes which represent the gravitational wave phenomena. The background gravitational waves in an expanding universe may be detected in future experiments. We address the reader to some references on this important topic, see Refs. \cite{grisha,shapiroo} and references therein.

At linear level, the computation of tensorial modes can be made by considering fluctuations around a given background solution, such that
\begin{eqnarray}
g_{\mu\nu} = g_{\mu\nu}^B + h_{\mu\nu},
\end{eqnarray}
where the superscript $B$ indicates the background metric and $|h_{\mu\nu}| << |g_{\mu\nu}^B|$ such that we can use the linear approximation.

The tensorial modes are represented by the traceless, divergence-free spatial components of $h_{\mu\nu}$ that behaves as pure tensorial quantity on the three-dimensional spatial section. That is,
the components $h_{ij}$ are such that $h_{kk} = \partial_{k}h_{ki} = 0$.
The tensorial modes are invariant by any coordinate transformation. In this sense, in their study, we can impose any coordinate condition or use the gauge-invariant formalism. For simplicity we use the synchronous coordinate condition, $h_{\mu0} = 0$. The perturbation of the unimodular condition implies,
\begin{eqnarray}
h_{kk} = 0,
\end{eqnarray}
which is already encoded in the condition to have pure tensorial modes.

Using the previous relations, i.e., the perturbation of the Ricci components, let us now perturb the Ricci scalar and the energy-momentum tensor. We restrict ourselves only to the tensorial modes. This implies that the only non zero components are the ones with the form \cite{wein-b},
\begin{eqnarray}
\delta R_{ij} &=& - \frac{\ddot h_{ij}}{2} + \frac{H}{2}\dot h_{ij} - 2H^2h_{ij},\\
\delta T_{ij} &=&  - p h_{ij}.
\end{eqnarray}
With the help of the background equations and after perform a Fourier decomposition, we find that the gravitational radiative modes obey the equation,
\begin{eqnarray}
\label{egw}
\ddot h_{ij} - H\dot h_{ij} - 2(\dot H + H^2)h_{ij} + \frac{k^2}{a^2}h_{ij} = 0.
\end{eqnarray}
This is exactly the equation for gravitational waves in GR with a perfect fluid, see Ref. \cite{wein-b}. This is by itself a surprising feature taking into account the many differences between GR and the nonconservative unimodular gravity. However, the evolution of gravitational waves
in the above described unimodular cosmological scenario and in the standard cosmological scenario should be very different: the expressions for the background functions $H$ and $\dot H$ are not the same, leading to special signatures in the
evolution of gravitational waves.

Let us return for the moment to equation (\ref{egw}) and review some of its property in the case the universe is dominated by radiation and a cosmological constant $\Lambda$. An analytic solution can be obtained when $\Lambda_{\rm U} = 0$. In this case, writing the scale factor in terms of the conformal time $\eta$ defined as $dt = ad\eta$, we obtain $a = a_0 \eta$, and the equation becomes,
\begin{eqnarray}
\label{gwe1}
h'' - 2\frac{h'}{\eta} + \biggr\{k^2 + \frac{2}{\eta^2}\biggl\}h = 0,
\end{eqnarray}
where we have, for simplicity, ignored the polarization indices. 
 This equation can be solved analytically for an equation of state
$p = \omega\rho$. The solution reads \cite{grishagw}
\begin{eqnarray}
\label{gws1}
h = \eta^{(5 + 3\omega)/[2(1 + 3\omega)]}\biggr\{A_1 J_\nu(k\eta) + A_2J_{-\nu}(k\eta)\biggl\}.
\end{eqnarray}
where $A_{1,2}$ are constants and $J_\nu(x)$ is the Bessel function of order $\nu$, with $\nu = 3(1 - \omega)/[2(1 + 3\omega)]$.

For a radiative universe ($\omega = 1/3$), the solution is
\begin{eqnarray}
\label{gws1}
h = \eta^{3/2}\biggr\{A_1J_{1/2}(k\eta) + A_2J_{-1/2}(k\eta)\biggl\}.
\end{eqnarray}
 The solution here is the same as in a cosmological model with only radiation in GR. Remember, however, that in the unimodular model the solution is valid irrespective of the choice made for $p$.

The solutions (\ref{gws1}) have two asymptotic behavior:
\begin{eqnarray}
\eta \rightarrow 0 \quad &\Rightarrow&\quad  h \propto (A_1\eta^3 + A_2\eta),\\
\eta \rightarrow \infty \quad &\Rightarrow& \quad h \propto \eta\cos(k\eta + \delta),
\end{eqnarray}
where $\delta$ is a phase.
Hence, initially the solutions exhibit growing modes which for large values of time becomes a growing oscillatory mode.

We remark {\it en passant} that for the case of the de Sitter solution ($\omega = - 1$, $\nu = 3/2$), the conformal time
range is $- \infty < \eta < 0$. Initially there are growing oscillations that asymptotically become either growing or decreasing modes, with no oscillations.

For the other possible solutions for the background in unimodular gravity (with $\Lambda_{\rm U} > 0$ or $\Lambda_{\rm U} < 0$), it seems not possible to obtain an closed expression for $h$. However, we can integrate numerically. First we do it in a qualitative way rewriting equation (\ref{egw}) for the scale factor.
The result is,
\begin{eqnarray}
\label{gw-eq}
h''  + \frac{\dot H}{H^2}\frac{h'}{a} + \biggr\{\frac{k^2}{H^2 a^4} - 2 \biggr[\frac{\dot H}{H^2} + 1\biggl]\frac{1}{a^2}\biggl\}h = 0.
\label{hnug}
\end{eqnarray} 
Now we need the expressions for $H$ and $\dot H$ in terms of the scale factor. In order to do so, we use (\ref{ce1}), which already provides $\dot H$ in terms of $a$ since we use that $\bar\rho \propto a^{-4}$. Then we multiply that equation by $H$ and integrate. The resulting set of equations are,
\begin{eqnarray}
\frac{\dot H}{H_0^2} &=& - \frac{3}{2}\frac{\bar\Omega_0}{a^4},\\
\frac{H^2}{H_0^2} &=& \frac{3}{4}\frac{\bar\Omega_0}{a^4} + 1 - \frac{3}{4}\bar\Omega_0,
\end{eqnarray}
where we have defined,
\begin{eqnarray}
\label{dp}
\bar\Omega_{0} = \frac{8\pi G\bar\rho_0}{3H_0^2},
\end{eqnarray}
and we have fixed $a(t = t_0) = 1$, where $t_0$ is the present time. Remark that the density parameter (\ref{dp}) is perfectly logical in a GR cosmology. But in unimodular gravity this definition may be seen as a choice since $\bar\rho$ contains also the pressure.

We must make now a more detailed and realistic comparison with the standard $\Lambda$CDM model in GR. This is the object of the next section.

\section{Comparing unimodular cosmology with the standard cosmological model}

Here, we will make a more detailed comparison of the unimodular gravity the evolution of the gravitational wave in the standard $\Lambda$CDM model.
 
In the $\Lambda$CDM model the matter content is given by radiation, pressureless matter and the cosmological constant.
In this case, the Friedmann equations read,
\begin{eqnarray}
\frac{H^2}{H^2_0} &=& \frac{\Omega_{r0}}{a^4} + \frac{\Omega_{m0}}{a^3} + \Omega_{\Lambda},\\
\frac{\dot H}{H_0^2} &=& - \frac{3}{2}\frac{\Omega_{m0}}{a^3}-2\frac{\Omega_{r0}}{a^4},
\end{eqnarray}
where the subscript $m$ indicates {\it matter}, the subscript $r$ refers to {\it radiation} and the density parameter is defined as before. The subscript $0$ indicates today's values of the parameters. The matter component scales as $a^{-3}$ while the radiation component scales as $a^{-4}$.
The constraint relation is now,
\begin{eqnarray}
1 = \Omega_{r0} + \Omega_{m0} +  \Omega_{\Lambda}.
\end{eqnarray}

Now, we go back to discuss the features of the nonconservative unimodular cosmological model. The main point, at this stage, is what we interpret as the matter parameter.
The equations in unimodular gravity are sensitive only to the combination $\rho + p$. Moreover, the system of equations is underdetermined, hence an ansatz must be introduced. Our choice exposed above leads to expressions that are similar to the radiative case in GR with a cosmological constant. This property lead us to define,
\begin{eqnarray}
\bar\rho = \rho + p = \frac{\bar\rho_0}{a^4},
\end{eqnarray}
with $\bar\rho_0$ indicating the value of the redefined density $\bar\rho$ today.
Using the definitions for the density parameter already shown in the previous section, we obtain equation (\ref{cec}) with solutions (\ref{sol1}-\ref{sol3}). The factor $3/4$ in these expressions is a consequence of our choices but allows us to represent the background expansion in a familiar form. Even if the
the unimodular equations are sensible only to the combination $\rho + p$, if we fix $p = \rho/3$ and using the same definitions for the density parameter, we obtain exactly the equations for a cosmology with radiation and a cosmological constant, as we could expect. But we will keep ourselves within the context of the unimodular structure with a generalized conservation law.
 
 With these definitions in mind, we will plot the cosmological evolution of gravitational waves in two scenarios: the unimodular model and the concordance $\Lambda$CDM model. We remark that in both cases the evolution of gravitational waves is dictated by equation (\ref{gw-eq}), differing only on the background functions $H$ and $\dot H$.
 
\begin{figure}[!t]
\includegraphics[width=0.45\textwidth]{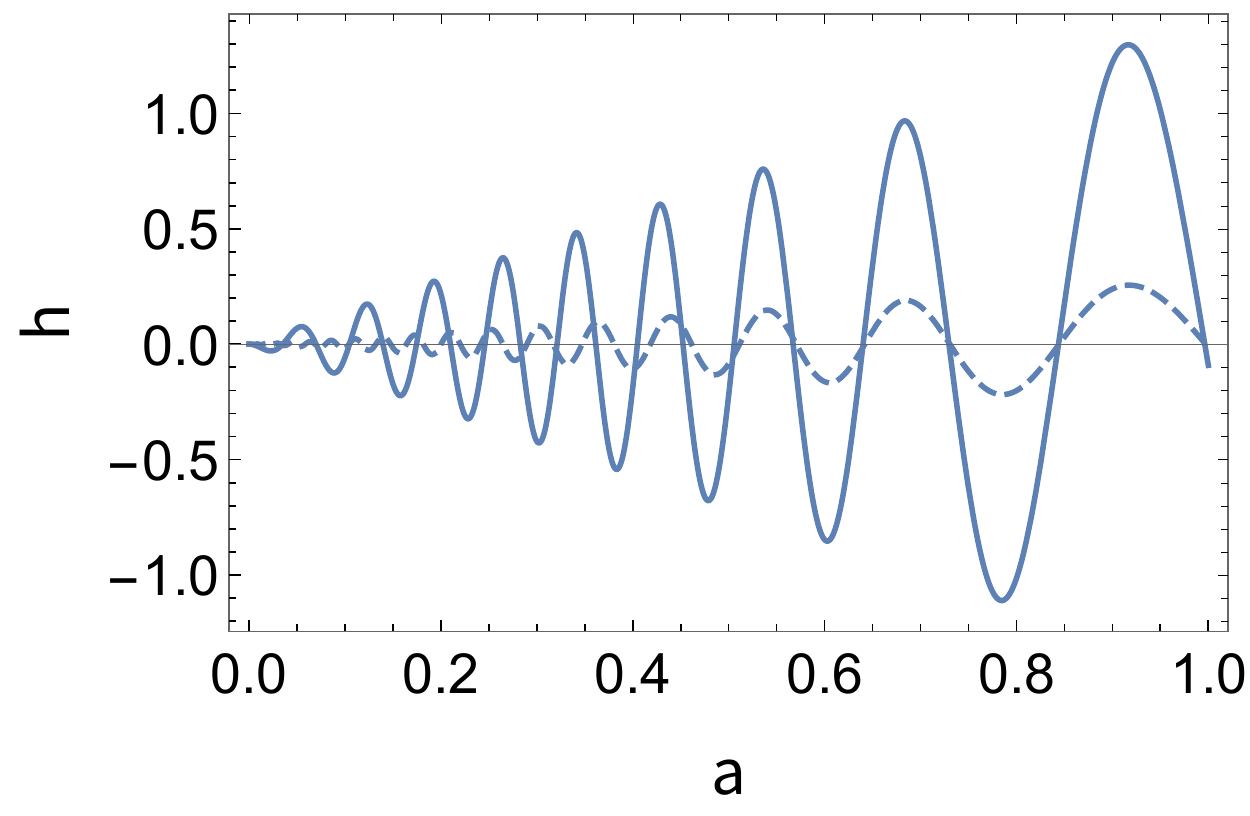}
\caption{Evolution of gravitational waves in unimodular gravity (continuous line) with $\bar\Omega_0 = 0.05$ and in the $\Lambda$ CDM model (dashed line) with $\Omega_{m0} = 0.3$ and $\Omega_{r0} = 5 \times 10^{-5}$. The initial conditions are the same.}
\label{fig2}
\end{figure}

\begin{figure}[!t]
\includegraphics[width=0.45\textwidth]{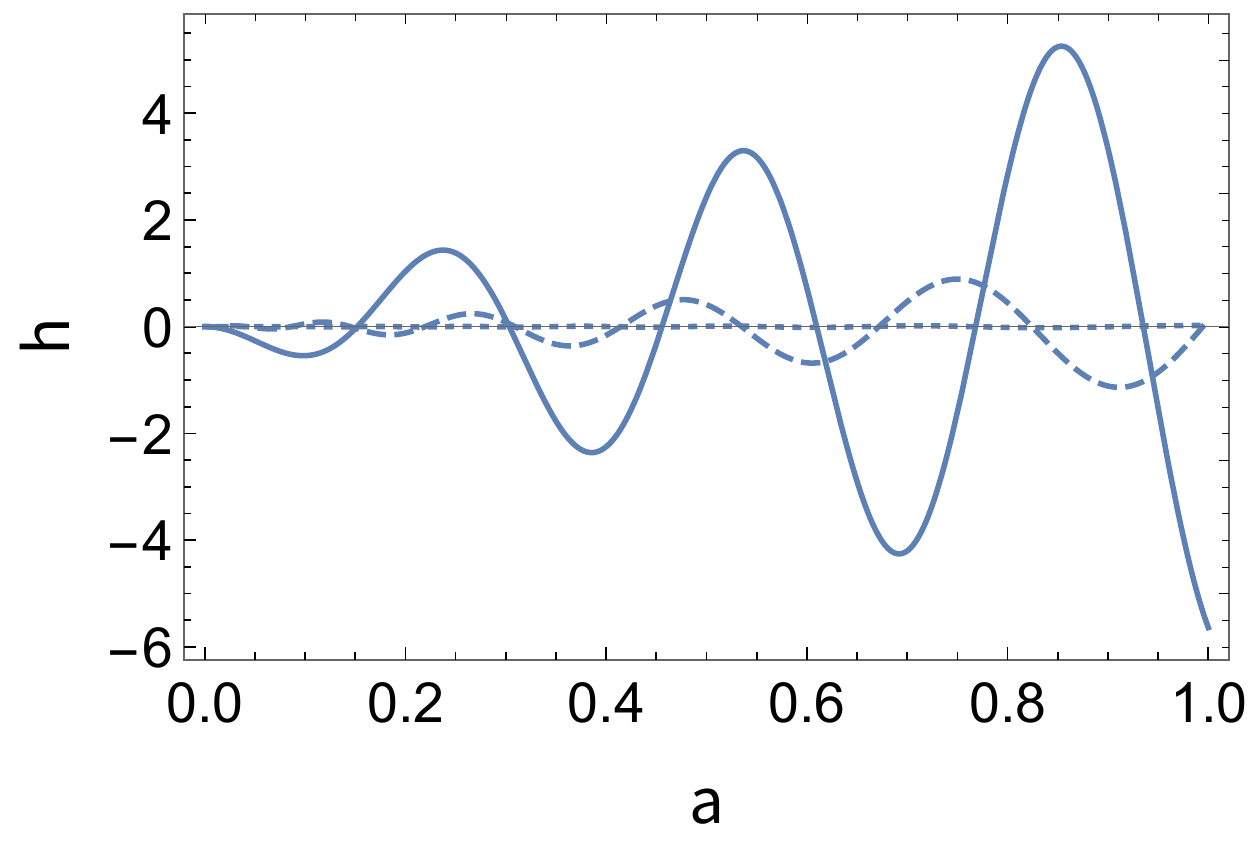}
\caption{Evolution of gravitational waves in unimodular gravity (continuous line) with $\bar\Omega_0 = 0.95$ and in the $\Lambda CDM$ model (dashed line) with $\Omega_{m0} = 0.3$ and $\Omega_{r0} = 5 \times 10^{-5}$. The initial conditions are the same.}
\label{fig3}
\end{figure}

First we fix $\bar\Omega_0 = 0.05$, corresponding to the total amount of baryonic matter, for the unimodular. For the $\Lambda$CDM model we fix $\Omega_{m0} = 0.3$ and $\Omega_{r0} = 5\times10^{-5}$ for the standard model. The result is displayed in Fig. \ref{fig2}. In Fig. \ref{fig3} we repeat the computation, but imposing now $\bar\Omega_0 = 0.95$, keeping the same values for the $\Lambda$CDM model, in order to stress the difference between the models. We remark that the unimodular case present a higher amplification of the gravitational waves
compared with the standard cosmological model for any value of the parameter $\Omega_0$. Also the frequency of the waves differ significantly.
This may have clear observational signatures. 
 
\section{Discussion and conclusions}

Unimodular gravity is a theory of gravity almost as old as GR. It is characterized by the condition
$g = $ constant. One of the attractive aspects of modular gravity is that by, imposing the usual conservation law for the energy-momentum tensor, the RG equations are recovered with an integration constant that can be identified with the cosmological term $\Lambda$. This fact brings a new perspective to the cosmological constant problem \cite{wein-a}. At quantum level, unimodular gravity has also many interesting features and discussed in \cite{quantum1}.

In this paper, we investigate one variant of unimodular gravity not very explored in the literature. As observed above, the conservation of the energy-momentum tensor is a free choice within the structure of theory, since the general invariance on diffeomorphism is lost when we fix the determinant of the metric. The consequences of not imposing the usual conservation laws were stressed here.
The first one is that, in the cosmological framework, the equations for the scale factor are sensitive only to the
combination $\rho + p$, irrespective of the fluid. Moreover, the equations are underdetermined, and they can not be solved without an ansatz. 

We have showed that the typical solution of radiation with (and without) a cosmological constant is consistently also a solution of the nonconservative unimodular gravity. As a remark, from this result, we can speculate that the cosmological constant is always {\it hidden} in the unimodular gravity, irrespective of the conservation of energy-momentum tensor or not.

In order to distinguish the unimodular cosmological model from the GR cosmological models, we have decided to investigate the evolution of gravitational waves in these different scenarios. The first obstacle is to identify the density parameters in unimodular gravity since all equations are only sensitive to the combination $\rho + p$, the entalphy of the system. This quantity scales as $a^{-4}$ as a legitime radiative field. We have identified this combination as the matter content of the model, and the
density parameter has been defined in order to keep close contact with the similar definitions in the standard cosmological model.

We have then studied the evolution of gravitational waves in the different cosmological scenarios emerging from
the unimodular gravity in comparison with the Standard Cosmological Model, with radiation, matter and cosmological constant. The main feature is that unimodular gravity as well as the radiation plus cosmological constant scenario in GR provides a stronger amplification of gravitational waves amplitudes compared the $\Lambda$CDM model.
Remark that if we choose a radiative equation of state, $p = \rho/3$, the unimodular scenario becomes identical to the radiative case with a cosmological constant in GR given the definition for the density parameter chosen.

Hence, it is possible in principle to distinguish the unimodular cosmological scenario from the $\Lambda$CDM model through the amplitude and the shape (frequency) of gravitational waves. However, in case future observations provide gravitational wave signals in agreement with the unimodular prediction, what could distinguish it from the radiative model with cosmological constant, since the features are very similar
at least in what concerns the evolution of gravitational wave? To try to answer this question, we must remember that structure formation in a cosmological model based in GR can not take place without a matter dominated phase, and we come back to the $\Lambda$CDM model. But, structure formation may take place in the unimodular scenario exhibited here, in spite of its radiative character, as some preliminary results indicate \cite{marcelo}. 

The results here reported are somehow qualitative, indicating the main characteristic features of the evolution of gravitational waves in unimodular gravity. A more detailed analysis of the observational signatures is necessary
to clearly distinguish the different scenarios. Of course, the analysis of structure formation, which is currently under study, may be another powerful way to investigate the viability of nonconservative version of unimodular cosmology.

However, the unimodular scenario discussed here seems to be incomplete because it is necessary an additional condition to solve the equations. In the present work we have imposed the same condition as
in reference \cite{trace}. Perhaps the nonconservative unimodular gravity must be complemented by some version of the 
holographic principle. We intend to perform in the future.

It would be interesting also to compare in details the results here obtained with those coming, for example, from the mimetic gravity.

\noindent
\section*{Acknowledgments}
 We thank to CNPq, CAPES, FAPES and PROPP/UFOP for financial support.

\end{document}